\documentclass[12pt]{article}

\usepackage{scicite}

\usepackage{times}

\usepackage{psfig}

\newenvironment{sciabstract}{%
\begin{quote} \bf}
{\end{quote}}

\def\HI{H\,{\sc i}}
\def\HII{H\,{\sc ii}}

\def\la{\ifmmode\stackrel{<}{_{\sim}}\else$\stackrel{<}{_{\sim}}$\fi} 
\def\ga{\ifmmode\stackrel{>}{_{\sim}}\else$\stackrel{>}{_{\sim}}$\fi} 

\newcounter{lastnote}
\newenvironment{scilastnote}{%
\setcounter{lastnote}{\value{enumiv}}%
\addtocounter{lastnote}{+1}%
\begin{list}%
{\arabic{lastnote}.}
{\setlength{\leftmargin}{.22in}}
{\setlength{\labelsep}{.5em}}}
{\end{list}}

\title{The Magnetic Field
of the Large Magellanic Cloud Revealed Through Faraday Rotation}

\author
{B. M. Gaensler,$^{1,2\ast}$, M. Haverkorn,$^{1}$ L. Staveley-Smith,$^{3}$
J. M. Dickey,$^{4}$ \\ N. M. McClure-Griffiths,$^{3}$
J. R. Dickel,$^{5}$ and M. Wolleben$^{6}$ \\
\\
\normalsize{$^{1}$Harvard-Smithsonian Center
for Astrophysics, 60 Garden Street MS-6, Cambridge, MA 02138, USA}\\
\normalsize{$^{2}$School of Physics, University of Sydney, NSW 2006,
Australia}\\
\normalsize{$^{3}$Australia Telescope National Facility, 
CSIRO, PO Box 76, Epping, NSW 1710, Australia}\\
\normalsize{$^{4}$Physics Department, University of Tasmania, GPO Box
252-21, Hobart, Tasmania 7001, Australia}\\
\normalsize{$^{5}$Astronomy Department, University of Illinois, 1002 
West Green Street, Urbana, IL 61801, USA}\\
\normalsize{$^{6}$Max-Planck-Institut f\"ur Radioastronomie, Auf dem
H\"ugel 69, D-53121 Bonn, Germany}\\
\normalsize{$^\ast$To whom correspondence should be addressed; E-mail:  
bgaensler@cfa.harvard.edu}
}

\date{}

\begin{document} 


\baselineskip24pt


\maketitle


\begin{sciabstract}
We have measured the Faraday rotation toward a large sample of polarized
radio sources behind the Large Magellanic Cloud (LMC), to determine
the structure of this galaxy's magnetic field. The magnetic field of
the LMC consists of a coherent axisymmetric spiral of field strength
${\bf \sim 1}$~microgauss.  Strong fluctuations in the magnetic field
are also seen, on small (${\bf <0.5}$~parsecs) and large (${\bf \sim
100}$~parsecs) scales.  The significant bursts of recent star formation
and supernova activity in the LMC argue against standard dynamo theory,
adding to the growing evidence for rapid field amplification in galaxies.
\end{sciabstract}


The Milky Way and many other spiral galaxies show well-organized,
large-scale magnetic fields\cite{bbm+96,bec00,hw02}, the existence of which
points to a powerful and ubiquitous process which organizes random motions
into coherent magnetized structures. The underlying mechanism is 
believed to be a dynamo, in which magnetic fields are slowly ordered
and amplified due to the interplay between turbulence and differential
rotation\cite{rss88,kul99}.
Magnetism in galaxies is usually mapped by observing the orientation of
polarized optical and radio emission from the galaxy itself\cite{hw02},
but these data have limited spatial resolution \cite{fbbs04} and can
be difficult to interpret\cite{sbs+98}. An alternative, direct
determination of the geometry and strength of
magnetic fields comes from the Faraday rotation of background
radio sources\cite{btj03,hbb98,gbf04}, an effect in which birefringence in
an intervening magneto-ionized source rotates the plane of linearly
polarized radiation.  Measurements of background rotation measures (RMs)
are free from the difficulties associated with studying polarized emission
produced by the source itself, which suffer from a complicated combination
of internal and external Faraday rotation, depolarization, and optical
extinction.  

We have studied the magnetic field of the LMC, using 1.4-GHz
polarization data recorded as part of a hydrogen line survey\cite{ksd+98}
carried out with the Australia Telescope Compact Array\cite{note2}.
Over a field of 130 square degrees, we have
calculated RMs for
291 polarized background sources.
Of this sample, about 100
sources lie directly behind the LMC.  We used 140 measurements of sources
lying outside the LMC to subtract a mean RM, presumably resulting
from foreground Faraday rotation in the Milky Way.  The resulting
distribution of residual Faraday rotation (Fig.~1) shows
a strong excess in  RM across the extent of the LMC. The
implied magnetic field demonstrates spatial coherence: the eastern
half of the galaxy shows predominantly positive RMs, while in the west
the RMs are mainly negative.

We have converted position on the sky to a location within the LMC
disk for each RM measurement, assuming that the galaxy is inclined to
the plane of the sky at an angle $i = 35^\circ$, with its line of
nodes at a position angle $\Theta = 123^\circ$ (measured north through
east)\cite{vdm04}. The resulting dependence of RM against position angle
within the disk of the LMC (Fig.~2) shows a 
systematic variation, with the maximum mean RM occurring near the
line of nodes.  These data can be well fit by a cosinusoid of amplitude
RM$_0 = 53\pm3$~rad~m$^{-2}$, demonstrating that the coherent component of
the LMC's magnetic field has an axisymmetric spiral geometry\cite{khb89},
as is seen in other galaxies\cite{bec00,hw02,khb89}. The phase
of this cosinusoid corresponds to the pitch angle, $p$, of the spiral
field\cite{khb89}, but in this case we can only infer a weak constraint,
$|p| \la 20^\circ$, especially given that the uncertainty on $\Theta$
is $\ga10^\circ$\cite{vdm04}.

In addition to the coherent field, a structure function analysis indicates
random fluctuations in RM, with a standard deviation $\Sigma_{\rm RM}
=  81$~rad~m$^{-2}$, and occurring on a characteristic angular scale
of $\approx 0.1^\circ$, or $L \approx 90$~parsecs at the distance to
the LMC of 50~kpc. This may represent the evolved supernova
remnants (SNRs) and wind bubbles
whose interlocking shells dominate the morphology of ionized gas in the
LMC on this scale\cite{mea80}.

To estimate the relative strength of the ordered and random field
components, we assumed that ionized gas in the LMC consists of a
disk of projected thickness $D$, in which cells of linear size $L$
contain clumps of ionized gas of filling factor $f$ and density
$n_e$\cite{note1}.
In each cell the magnetic field is comprised of a uniform component of
strength $B_0$ plus a randomly oriented component of strength $B_R$.
If $B_R$ is uncorrelated with fluctuations in $n_e$, it can be shown that:
\begin{equation}
\frac{\Sigma_{\rm RM}}{|{\rm RM}_0|} \approx \left[ \frac{L}{2fD} 
\left( 1 + \frac{2}{3}\frac{B_R^2}{B_0^2 \sin^2 i} \right) \right]^{1/2}
\label{eqn_sigrm1}
\end{equation}
The occupation length, $fD$, of ionized gas is ${\rm DM}_0^2/{\rm EM}_0
\approx 530$~pc, where DM$_0 \equiv \int n_e dl \approx 100$~cm$^{-3}$~pc
and EM$_0 \equiv  \int n_e^2 dl \approx 19$~pc~cm$^{-6}$ are the
average dispersion measure (DM) and median extinction-corrected emission
measure (EM) integrated through the LMC, respectively\cite{note2}.
Equation~(\ref{eqn_sigrm1}) then implies $B_R/B_0 =3.6$: the
random field dominates the ordered field, as seen in many other
galaxies\cite{bec00,bssw03}.

The actual values of $B_0$ and $B_R$ can also be estimated.  If $B_R$
and fluctuations in $n_e$ are uncorrelated, then the strength of the
ordered component of the LMC's magnetic field is
\begin{equation}
B_0 = \frac{|{\rm RM}_0|}{K~{\rm DM}_0 \sin i} \approx 1.1~\mu{\rm G},
\end{equation}
where $K = 0.81$~rad~m$^{-2}$~pc$^{-1}$~cm$^3$~$\mu$G$^{-1}$.
The strength of the random field is then $B_R = 3.6 B_0 \approx 4.1$~$\mu$G,
and the total magnetic field strength
on large scales is $B_T = (B_0^2 + B_R^2)^{1/2} \approx 4.3$~$\mu$G.
We note that in selected regions where
$B$ and $n_e$ are correlated, as might
result from compression in SNR shocks,
the above approach overestimates $B_0$ and underestimates $B_R$, each
by factors of $\sim2$ \cite{bssw03}.

The polarized background
sources are not randomly distributed across the LMC's extent, but tend
to avoid areas of bright H$\alpha$
emission (Fig.~1). Specifically, 
background sources are increasingly depolarized as they propagate
through regions of higher EM (Fig.~3).
This suggests that we are
observing beam depolarization, in which small-scale fluctuations in
foreground RM produce interference between polarized rays along adjacent
sightlines\cite{sbs+98}. Beam depolarization is only significant when
the angular scale of RM fluctuations is smaller than
the resolution of the data and the scale of intrinsic polarized
structures. While the resolution here, $\sim40$~arcsec, is comparatively
large, extragalactic sources are typically of much smaller angular extent:
for 1.4-GHz flux densities in the range 10--300~millijanskys as observed
here, the median angular size is only $\approx6$~arcsec\cite{wfpl93}, or
$\approx 1.5$~parsecs when projected against the LMC. The depolarization
(Fig.~3) implies strong RM fluctuations on
scales $l \ll 1.5$~parsecs.  In such a situation, beam depolarization
reduces the intrinsic linearly polarized intensity, $P_0$, to a
level\cite{bur66,sbs+98}:
\begin{equation}
P = P_0 e^{-2\sigma^2_{\rm RM} \lambda^4},
\label{eqn_burn}
\end{equation}
where $\lambda$ is the observing wavelength and $\sigma_{\rm RM}$
is the standard deviation in RMs across the source.
To account for the dependence of $P/P_0$ on EM
(Fig.~3), we need to relate $\sigma_{\rm RM}$ to the EM
along a given sightline. If the magnetic field is uncorrelated with the
ionized gas density, we expect that $\sigma_{\rm RM} = k~{\rm EM}^{1/2}$,
where $k$ is a constant.  With this assumption, the fluctuating
magnetic field on a scale $l$~parsecs needed to produce the observed
depolarization has a strength $B_r \approx k (lK^2/3)^{-1/2}$~$\mu$G.
In Figure~3, Equation~(\ref{eqn_burn})
has been fit to the data for $P_0 \approx 0.104$ and $k \approx
1.8$~rad~m$^{-2}$~pc$^{-1/2}$~cm$^3$.  Assuming $l < 0.5$~parsecs, we
find that $B_r > 5$~$\mu$G. We thus infer that there are significant RM
and magnetic field fluctuations on sub-parcsec scales in the ionized
gas of the LMC. This
phenomenon is also seen in our own Galaxy, and may trace the turbulent
winds and \HII\ regions of individual stars\cite{lea87,hgm+04}.

Most spiral galaxies are long-lived systems that exhibit significant
rotational shear and that experience relatively constant star-formation
rates over long periods of time. Coherent magnetic fields in these
galaxies are believed to be produced by a dynamo mechanism, in which
small-scale turbulent magnetic fields are amplified and ordered by
cyclonic motions and differential rotation\cite{bec00,rss88,kul99}.
However, in galaxies dominated by sudden bursts of star formation and
supernova activity, the dramatic injection of energy should disrupt the
slow monotonic increase of the large-scale field produced by a standard
turbulent dynamo\cite{kro94,cb04}. The LMC has experienced several
intense bursts of star formation over the past $\sim4$~Gyr triggered
by repeated close encounters with the Milky Way and with the Small
Magellanic Cloud\cite{osm96,bc05}, and yet still maintains a coherent
spiral magnetic field. Combined with previous results demonstrating
the presence of ordered magnetic fields in young galaxies for
which the dynamo has had little time to operate\cite{kpz92}, and in
irregular galaxies which lack significant amounts of 
rotation\cite{cbk+00}, there is now evidence that standard dynamo
processes are ineffective in the LMC and these other galaxies.
There are several viable alternatives to explain the coherent magnetic
fields that we observe.  Potentially most pertinent for the LMC is the
cosmic-ray driven dynamo, in which recent supernova activity generates
a significant population of relativistic particles.  The buoyancy of
these particles inflates magnetic loops out of the disk; adjacent loops
reconnect, and then are amplified by differential rotation to generate
a large-scale spiral field\cite{mss99,hkol04}. This mechanism not only
requires vigorous star formation, as has occurred recently for the LMC,
but has a time scale for amplification of only $\sim0.2$~Gyr\cite{hkol04},
and so can quickly generate large-scale magnetic fields before they are
dissipated by outflows and tidal interactions.  This process can thus
potentially account for the coherent fields seen in the LMC and other
galaxies\cite{ocsv00}.



\begin{thebibliography}{10}
\bibitem{note2}
Materials and methods are available as supporting material on Science
Online.
\bibitem{note1}
The depolarization demonstrated in Figure~3 implies
that the observed RMs do not probe ionized gas in bright individual
\HII\ regions, making this a reasonable assumption.
\end{thebibliography}

\begin{scilastnote}
\item We are grateful to Sungeun Kim for carrying out the original ATCA
observations which made this project possible.  We also thank Rainer
Beck, Richard Crutcher, Katarzyna Otmianowska-Mazur, Detlef Elstner
and Dmitry Sokoloff for useful discussions.  The Southern H-Alpha Sky
Survey Atlas (SHASSA) is supported by the National Science Foundation.
The Australia Telescope is funded by the Commonwealth of Australia for
operation as a National Facility managed by CSIRO.  B.M.G. acknowledges
the support of the National Science Foundation through grant AST-0307358,
and of the University of Sydney through the Denison Fund.
\end{scilastnote}

\vspace{5mm}
\noindent
{\bf Supporting Online Material} \\
www.sciencemag.org \\
Materials and Methods


\clearpage

\centerline{   }

\vspace{0.5cm}
\centerline{\bf FIGURE CAPTIONS}

\vspace{0.5cm}

\noindent {\bf FIGURE 1:} Faraday rotation measures through the Large
Magellanic Cloud.  The image shows the distribution of emission measure
toward the LMC in units of pc~cm$^{-6}$, derived from the Southern
H-Alpha Sky Survey Atlas \cite{gmrv01}.  The symbols show the
position, sign and magnitude of the baseline-subtracted RM measurements
\cite{note2}: filled and open circles (both marked in green)
correspond to positive and negative RMs, respectively, while asterisks
(marked in purple) indicate RMs which are consistent with zero within
their errors.  The diameter of each circle is proportional to the
magnitude of the RM, the largest positive and negative RMs being
$+247\pm13$~rad~m$^{-2}$ and $-215\pm32$~rad~m$^{-2}$, respectively.

\vspace{0.5cm}



\noindent {\bf FIGURE 2:} RM against position angle within the LMC. The
six data points are a binned representation of the 93 RMs which lie
within a radius of 3.5~degrees of the center of the EM distribution
seen in Fig.~1 (i.e., Right Ascension [J2000] $05^{\rm
h}16^{\rm m}03^{\rm s}$, Declination [J2000] $-68^\circ41'45''$),
plotted against deprojected position angle within the LMC, measured
from the line of nodes. The uncertainty on each datum is the weighted
standard error in the mean for RMs in that bin. The dashed line shows a
cosinusoidal least-squares fit to the unbinned data, with an amplitude of
$+53\pm3$~rad~m$^{-2}$ and an offset from zero of $+9\pm2$~rad~m$^{-2}$.
The phase of the cosinusoid is only weakly constrained, falling between
$\pm15^\circ$.  The fit is not a strong function of the center adopted
for the LMC.

\vspace{0.5cm}

\noindent {\bf FIGURE 3:} Polarized fraction of 81 background sources
as a function of EM (12 of the 93 sources shown in Fig.~2
have been excluded: six with EM~$>100$~pc~cm$^{-6}$, and six with an
observed EM~$\le 0$~pc~cm$^{-6}$ due to imperfect star subtraction).
A Galactic foreground contribution of 3~pc~cm$^{-6}$ has been subtracted
from each EM measurement. The uncertainty for each binned data-point
corresponds to the weighted standard error in the mean for each bin.
The observed depolarization as a function of EM cannot be a result of
source confusion or other observational selection effects, since sources
with RMs were identified from an image of linear polarization, in which
the weak signals from diffuse polarized emission show no correlation with
H$\alpha$ emission. It also cannot be due to excessive Faraday rotation
across our observing band (bandwidth depolarization), since for the
narrow frequency channels (8~MHz) used here, this effect would manifest
itself only for $|{\rm RM}| > 4000$~rad~m$^{-2}$, $\sim20$ times larger
than any RMs observed.  The dashed line shows a least-squares fit of
Equation~(\ref{eqn_burn}) to the unbinned data, assuming $\sigma_{\rm RM}
\propto {\rm EM}^{1/2}$.


\clearpage

\begin{figure}
\centerline{\psfig{file=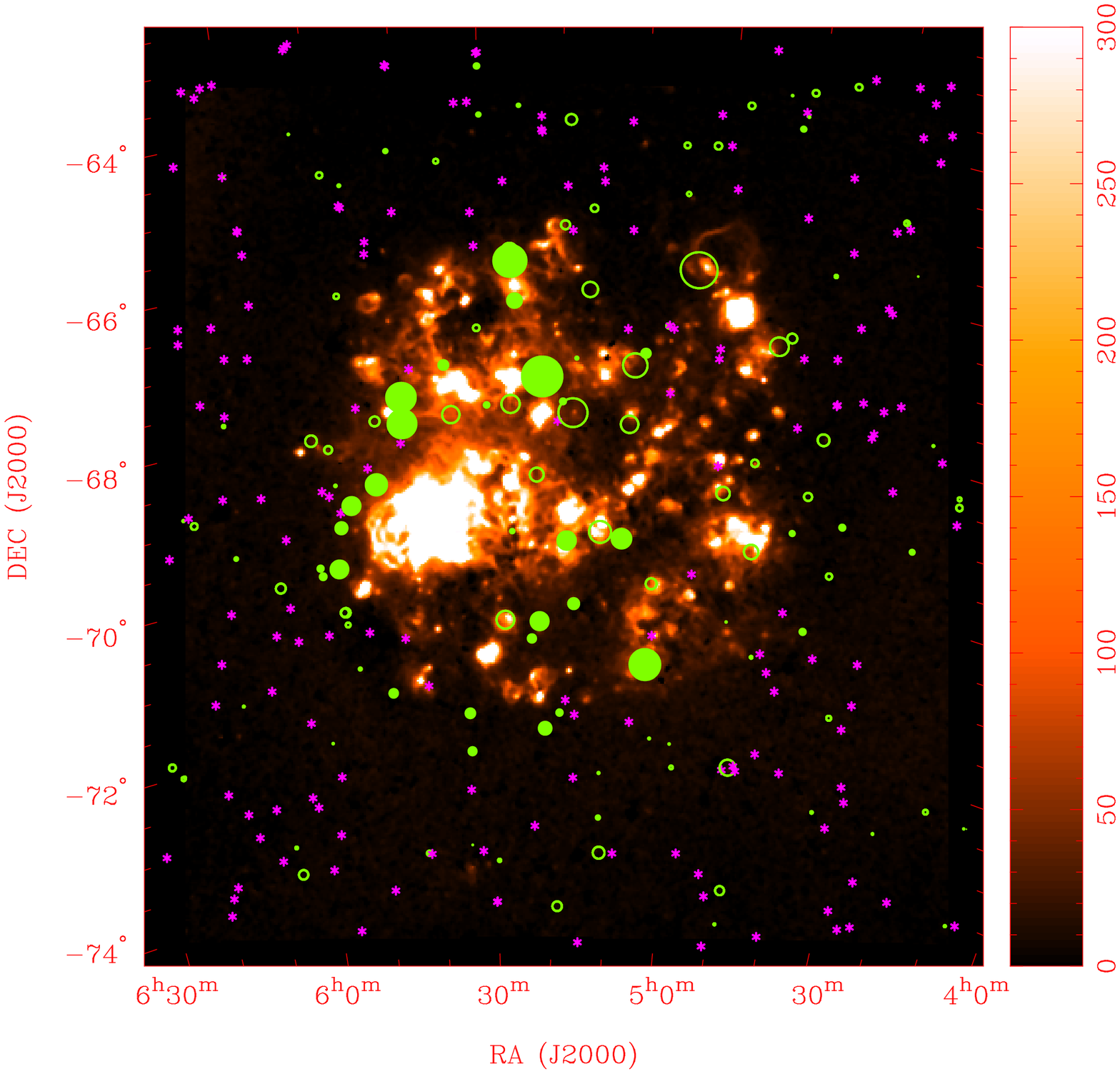,width=\textwidth,angle=270}}
\caption{}
\label{fig_rms}
\end{figure}

\begin{figure}
\centerline{\psfig{file=fig_rm_vs_pa.eps,width=\textwidth,angle=270}}
\caption{}
\label{fig_pa}
\end{figure}

\clearpage


\begin{figure}
\centerline{\psfig{file=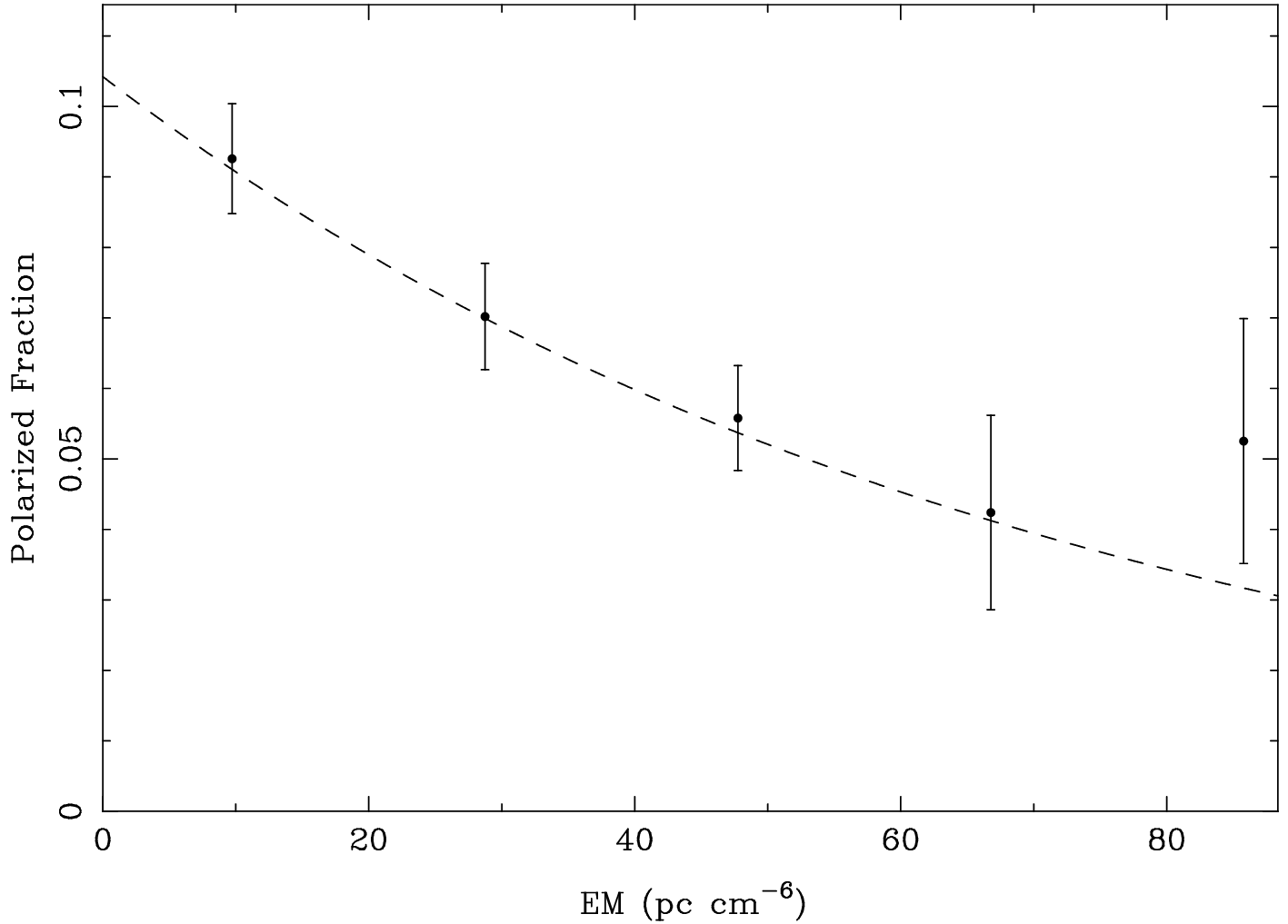,width=\textwidth,angle=270}}
\caption{}
\label{fig_em}
\end{figure}

\clearpage

\centerline{   }

\vspace{0.5cm}
\centerline{\bf SUPPORTING ON-LINE MATERIAL}
\vspace{0.5cm}

\centerline{\bf Materials and methods}

The RMs presented here were determined from a 1.4-GHz mosaic of 1300
pointings toward the LMC, made using the Australia Telescope Compact
Array over the period 1994~October to 1996~February at a resolution
of 40~arcsec (Kim et al, 1998, {\em The Astrophysical Journal}, v503,
p674).  Full polarization was recorded in this survey in 14 successive
channels of bandwidth 8~MHz each, with center frequencies between
1328 and 1432~MHz. Antenna gains were calibrated using observations
every $\sim30$~min of one of PKS~B0407--68 or PKS~B0454--810, while
absolute flux densities and polarization leakages were determined using
observations of PKS~B1934--638.  For each frequency channel, images were
formed in Stokes~$Q$ and $U$ at a pixel size of $13\times13$~arcsec$^2$,
were deconvolved using a maximum entropy algorithm, corrected for
primary beam attenuation, and then were combined and debiased to form
an image of linear polarization, $\mathcal{L}$~$=(Q^2 + U^2)^{1/2}$.
Using a  false-discovery rate algorithm (Hopkins et al., 2002, {\em
The Astronomical Journal}, v123, p1086) on an average of the 14 images
of $\mathcal{L}$, 324 polarized sources were identified which were
unresolved, did not correspond to a catalogued pulsar or supernova remnant
within the LMC, and had a fractional linear polarization between 0.3 and
50 per cent.  For each such source, values of Stokes~$Q$ and $U$ were then
extracted from the individual channel maps for nine pixels surrounding the
source peak. For each pixel, the RM and its uncertainty were determined
from the 14 pairs of $Q$ and $U$ values using the algorithm of Brown
et al, 2003, {\em The Astrophysical Journal Supplemental Series}, v145,
p213. An individual pixel's RM measurement was accepted only if the pixel
had a debiased signal-to-noise in polarization of $>1$ in at least eight
frequency channels, and if the quality of the RM fit was better than
90\%. A source's overall RM was considered valid if at least five of
nine pixels had acceptable RM fits, and if the dispersion in RM between
pixels was less than twice the average error in each pixel's RM. Of 324
polarized sources, 291 passed all these tests, with typical errors in
RM for each source of $\pm20$~rad~m$^{-2}$.  By considering 140 RMs
at a radius of $>4.5$~degrees from the LMC's center, we found that
a positive offset in RM was present due to foreground and background
contributions; a mean baseline was therefore subtracted from the data
in four separate quadrants: $+33.5\pm3.3$~rad~m$^{-2}$ in the southeast,
$+22.5\pm3.8$~rad~m$^{-2}$ in the southwest, $+49.0\pm4.5$~rad~m$^{-2}$
in the northwest, and $+31.3\pm3.4$~rad~m$^{-2}$ in the northeast.

To estimate the total dispersion measure through the LMC, we used
the fact that dispersion of the pulsed signals from the five
known radio pulsars in the LMC implies a mean total dispersion
measure DM~$\sim100$~cm$^{-3}$~pc (Crawford et al., 2001, {\em The
Astrophysical Journal}, v553, p367). The foreground contribution to the
DM from the Milky Way is $\approx50$~cm$^{-3}$~pc (Cordes \& Lazio,
preprint, http://arxiv.org/abs/astro-ph/0207156).  If the mean DM of
LMC pulsars corresponds to a location half way through the LMC disk,
then DM$_0$~$\approx 100$~cm$^{-3}$~pc.

Emission measures have been derived from the Southern H-alpha Sky Survey
Atlas (Gaustad et al., 2001, {\em Publications of the Astronomical Society
of the Pacific}, v113, p1326), by correcting smoothed and star-subtracted
H$\alpha$ emission in this region for both foreground extinction in the
Milky Way and internal extinction in the LMC.  In both cases, extinction
corrections were derived from \HI\ column density maps (Staveley-Smith
et al., 2003, {\em Monthly Notices of the Royal Astronomical Society},
v339, p87; Kim et al., 2003, {\em The Astrophysical Journal Supplemental
Series}, v148, 473) using typical gas-to-dust ratios for our Galaxy and
for the LMC (Pei, 1992, {\em The Astrophysical Journal}, v395, p130),
and assuming that extinction within the LMC is caused by only half of
the LMC \HI\ column.  H$\alpha$ intensity was then converted into EM units
assuming an electron temperature of 8000~K.

\end{document}